\documentclass[conference]{IEEEtran}
\IEEEoverridecommandlockouts
\usepackage{cite}
\usepackage{amsmath,amssymb,amsfonts}
\usepackage{algorithmic}
\usepackage{graphicx}
\usepackage{textcomp}
\usepackage{xcolor}
\newcommand{\mybibliography}{\bibliography{jour_short,conf_short,ref.bib}}

\usepackage{titlesec}
\let\subparagraph\relax
\titlespacing{\section}{0pt}{6pt plus 2pt minus 1pt}{4pt plus 1pt minus 1pt} 
\titlespacing{\subsection}{0pt}{4pt plus 2pt minus 1pt}{2pt plus 1pt minus 1pt} 
\usepackage{comment}

\DeclareMathOperator{\pa}{pa}

\usepackage{tikz}
\usetikzlibrary{bayesnet}
\usepackage{standalone}

\usepackage{physics}
\usepackage{multirow}
\usepackage{tablefootnote}
\usepackage{amssymb}

\usepackage{subfigure}
\usepackage{multirow}
\usetikzlibrary{tikzmark}
\usepackage{booktabs}
\usepackage{tabularx}
\newcolumntype{s}{>{\hsize=.7\hsize}X}

\usepackage[nolist,nohyperlinks]{acronym}
\acrodef{GCS}{global coordinate system}
\acrodef{LCS}{local coordinate system}
\acrodef{BS}{base station}
\acrodef{MS}{mobile station}
\acrodef{RIS}{reconfigurable intelligent surface}
\acrodef{OFDM}{orthogonal frequency division multiplexing}
\acrodef{UPA}{uniform planar array}
\acrodef{MIMO}{multiple-input multiple-output}
\acrodef{AWGN}{additive white Gaussian noise}
\acrodef{DAG}{directed acyclic graph}
\acrodef{PDF}{probability density function}
\acrodef{MCMC}{Markov chain Monte Carlo}
\acrodef{NUTS}{No-U-Turn Sampler}
\acrodef{CE}{channel estimation}
\acrodef{LOS}{line-of-sight}
\acrodef{ULA}{uniform linear array}
\acrodef{CRLB}{Cramér–Rao lower bound}
\acrodef{PEB}{position error bound}
\acrodef{HRIS}{hybrid reconfigurable intelligent surface}
\acrodef{REB}{rotation error bound}
\acrodef{NLOS}{non-line-of-sight}
\acrodef{TOA}{time of arrival}
\acrodef{TDOA}{time difference of arrival}
\acrodef{AOA}{angle of arrival}
\acrodef{UE}{user equipment}
\acrodef{FIM}{fisher information matrix}
\acrodef{NLS}{non-linear least squares}
\acrodef{ML}{maximum likelihood}
\acrodef{RMSE}{root mean square error}
\acrodef{CDF}{cumulative distribution function}
\usepackage{etoolbox}
\makeatletter
\patchcmd{\@maketitle}
  {\addvspace{0.5\baselineskip}\egroup}
  {\addvspace{-1\baselineskip}\egroup}
  {}
  {}
\makeatother
\begin{document}
\title{Positioning via Probabilistic Graphical Models in RIS-Aided Systems with Channel Estimation Errors\\
\author{\IEEEauthorblockN{Leonardo~Terças\IEEEauthorrefmark{1} and Markku~Juntti\IEEEauthorrefmark{1}}
\IEEEauthorblockA{
    \IEEEauthorrefmark{1}Centre for Wireless Communications (CWC), University of Oulu, Finland}

\IEEEauthorblockA{E-mail: \{leonardo.tercas,  markku.juntti\}@oulu.fi}}
}
\maketitle
\begin{abstract}
We propose a 6D Bayesian-based localization framework to estimate the position and rotation angles of a mobile station (MS) within an indoor reconfigurable intelligent surface (RIS)-aided system. This framework relies on a probabilistic graphical model to represent the joint probability distribution of random variables through their conditional dependencies and employs the No-U-Turn Sampler (NUTS) to approximate the posterior distribution based on the estimated channel parameters. Our framework estimates both the position and rotation of the mobile station (MS), in the presence of channel parameter estimation errors. We derive the Cramér–Rao lower bound (CRLB) for the proposed scenario and use it to evaluate the system's position error bound (PEB) and rotation error bound (REB). We compare the system performances with and without RIS. The results demonstrate that the RIS can enhance the positioning accuracy significantly.
\end{abstract}

\begin{IEEEkeywords}
Bayesian, RIS, graphical models.
\end{IEEEkeywords}
\section{Introduction}
\Ac{RIS} have been shown in recent studies to significantly enhance the performance of communications, localization, and sensing systems \cite{liu2021reconfigurable, chen2022reconfigurable}. \ac{RIS} consists of a two-dimensional array of numerous small, low-cost, reconfigurable elements capable of manipulating electromagnetic waves \cite{wu2019towards}. By adjusting the phase shift, amplitude, polarization, and frequency of these elements, \ac{RIS} can control signal propagation, improving signal quality, coverage, and energy efficiency. This technology has been extensively studied and is expected to be a key enabler for improving communication and sensing in next-generation wireless networks \cite{behravan2022positioning}. \ac{RIS} is particularly important for indoor applications, where it holds great potential to enhance communications quality, improve localization accuracy, and mitigate multipath effects by shaping signal propagation in real time \cite{zheng2022survey,bayraktar2022multidimensional}.

Among the many studies on \ac{RIS}, \ac{RIS}-aided or \ac{RIS}-enabled positioning stands out. In these studies, \ac{RIS} acts as an anchor with a known state (i.e., location and orientation) and exploits the reflected signals from the surface to improve or assist \ac{MS} localization. However, implementing indoor \ac{RIS}-aided mmWave \ac{MIMO} systems faces several challenges, including managing multipath reception, obstructed \ac{LOS}, and the high density of scatterers and clutter. Channel estimation is particularly complex in mmWave systems due to their susceptibility to blockage and attenuation, making it difficult to provide accurate feedback for optimal beamforming and signal reflection in dynamic indoor environments \cite{wei2021channel, zheng2022survey}. Many works have derived the \ac{CRLB} and the \ac{PEB} for two- and three-dimensional localization problems in scenarios containing \ac{RIS} \cite{ghazalian2024joint,liu2022cramer}. 

The potential of RIS-aided mmWave systems to enhance localization capabilities alongside communications by utilizing sparse reconstruction algorithms for high-resolution channel estimation in 3D indoor environments has been studied in \cite{bayraktar2022multidimensional}. A joint 3D localization of a hybrid reconfigurable intelligent surface and a user, utilizing a multistage approach for 6D parameter estimation (position and rotation) has also been explored in \cite{ghazalian2024joint}, along with the derivation of the \ac{CRLB} for the system. In \cite{liu2022cramer}, Liu \textit{et al.} investigate the \ac{CRLB} for location estimation error in multiple-RIS-aided mmWave positioning systems, focusing on the \ac{PEB} and \ac{REB}. Zhang \textit{et al.} \cite{zhang2023multi} propose a single-base-station localization method for multiple \acs{RIS} in \ac{NLOS} 3D environments by estimating the position of the user using time of arrival, angle of arrival, and time difference of arrival.

Conventional optimization-based algorithms like non-linear least squares and maximum likelihood estimators \cite{ghazalian2024joint, liu2022cramer, zhang2023multi} have been widely used for user location estimation, but they often require extensive measurements or iterations to achieve accuracy, which may not be feasible in some applications. In contrast, Bayesian inference, using probabilistic graphical models, offers an alternative by representing complex relationships through probability rules and incorporating uncertainties and prior knowledge. This method estimates position by sampling from the joint posterior distribution, providing a distinct approach to target positioning. In this paper, we propose a 6D Bayesian-based localization framework to estimate both the position and rotation of a \ac{MS} in an RIS-aided indoor environment. Our main contribution is the development of this framework, which integrates Bayesian inference with probabilistic graphical models to improve location and orientation accuracy. Additionally, we perform a comparative analysis with scenarios that exclude \ac{RIS}, focusing on the performance gains in both position and rotation estimation introduced by the presence of RIS. To assess the proposed framework’s performance, we analyze the \ac{CDF} of the estimation error, along with the \ac{PEB} and \ac{REB} derived from the \ac{CRLB}. 

\section{System Model}
\label{sec:system}
\subsection{Coordinate System}
The coordinate system considered in this work is defined in \cite{3gpp_38.901_V.16.1.0}. The Cartesian representation of a point $(x, y, z)$ on the unit sphere are determined by its spherical coordinates described by $(\rho = 1, \theta, \phi)$, where $\rho$ represents the unit radius, $\theta$ denotes the zenith angle measured from the positive $z$-axis, and $\phi$ indicates the azimuth angle measured from the positive $x$-axis within the $x$-$y$ plane. The Cartesian coordinates are expressed as follows
\begin{equation}
    \scalebox{0.85}{$
    \setlength{\arraycolsep}{2pt}
    \medmuskip = 1mu
    \hat{\boldsymbol{\rho}} =
    \begin{bmatrix}
        x \\
        y \\
        z 
    \end{bmatrix} 
    = 
    \begin{bmatrix}
        \rho \sin( \theta) \cos( \phi)\\
        \rho \sin( \theta) \sin( \phi) \\
        \rho \cos( \theta)
    \end{bmatrix}, 
    $}
\end{equation}
\noindent {where this coordinate point represents a location in the \ac{GCS}, the spherical coordinates can be computed as $\theta~=~\arccos(z)$ and $\phi~=~\arctan2( y, x )$\footnote{The $\arctan$2 function computes the angle between the positive $x$-axis and the point $(x, y)$ in the Cartesian plane. The result is in the range $[-\pi, \pi]$.}.} Considering an arbitrary 3D rotation, it is possible to compute its corresponding position in the \ac{LCS}. The composite rotation matrix representing any arbitrary 3D rotation is given by \eqref{eq:rot_matrix}
\begin{equation}
    \mathbf{R} = \mathbf{R}_\text{z}(\alpha)\mathbf{R}_\text{y}(\beta)\mathbf{R}_\text{x}(\gamma),
    \label{eq:rot_matrix}
\end{equation}
where $\alpha$ is the bearing angle, $\beta$ is the downtilt angle, and $\gamma$ is the slant angle. The matrices $\mathbf{R}_\text{z}(\alpha)$, $\mathbf{R}_\text{y}(\beta)$, and $\mathbf{R}_\text{x}(\gamma)$ represent the rotations around the $z$-axis, $y$-axis, and $x$-axis, respectively, and are given by
\begin{equation}
    \scalebox{0.85}{$
    \setlength{\arraycolsep}{2pt}
    \medmuskip = 1mu
    \mathbf{R}_\text{z}(\alpha) =
    \begin{bmatrix}
        \cos(\alpha) & -\sin(\alpha) & 0 \\
        \sin(\alpha) & \cos(\alpha)  & 0 \\
        0 & 0 & 1
    \end{bmatrix}, 
    \quad
    \mathbf{R}_\text{y}(\beta) =
    \begin{bmatrix}
        \cos(\beta) & 0 & \sin(\beta) \\
        0 & 1  & 0 \\
        -\sin(\beta) & 0 & \cos(\beta)
    \end{bmatrix},
    $}
    \nonumber
    \label{eq:rotation_rz_ry}
\end{equation}

\begin{equation}
    \scalebox{0.85}{$
    \setlength{\arraycolsep}{2pt}
    \medmuskip = 1mu
    \mathbf{R}_\text{x}(\gamma) =
    \begin{bmatrix}
        1 & 0 & 0 \\
        0 & \cos(\gamma)  & -\sin(\gamma) \\
        0 & \sin(\gamma) & \cos(\gamma)
    \end{bmatrix}.
    $}
    \label{eq:rotation_rx}
\end{equation}

The relationship between the spherical angles $(\theta, \phi)$ in the \ac{GCS} and the spherical angles $(\theta', \phi')$ in the \ac{LCS}, considering the rotation defined by \eqref{eq:rot_matrix}, are expressed as follows
\begin{equation}
    \scalebox{0.75}{$
    \setlength{\arraycolsep}{2pt}
    \medmuskip = 0.9mu
    \theta' = \arccos \left( \begin{bmatrix}
        0 \\
        0 \\
        1 
    \end{bmatrix}^\text{T}  \mathbf{R}^{T} \hat{\boldsymbol{\rho}} \right), 
    \quad
    \phi' = \arctan2 \left( \begin{bmatrix}
        0 \\
        1 \\
        0 
    \end{bmatrix}^\text{T}  \mathbf{R}^{T} \hat{\boldsymbol{\rho}},  \begin{bmatrix}
        1 \\
        0 \\
        0 
    \end{bmatrix}^\text{T}  \mathbf{R}^{T} \hat{\boldsymbol{\rho}} \right).
    $}
    \label{eq:theta_phi'}
\end{equation}

\subsection{Signal Model}
\begin{figure}[!t]
    \centering
    \includegraphics[width=0.9\columnwidth]{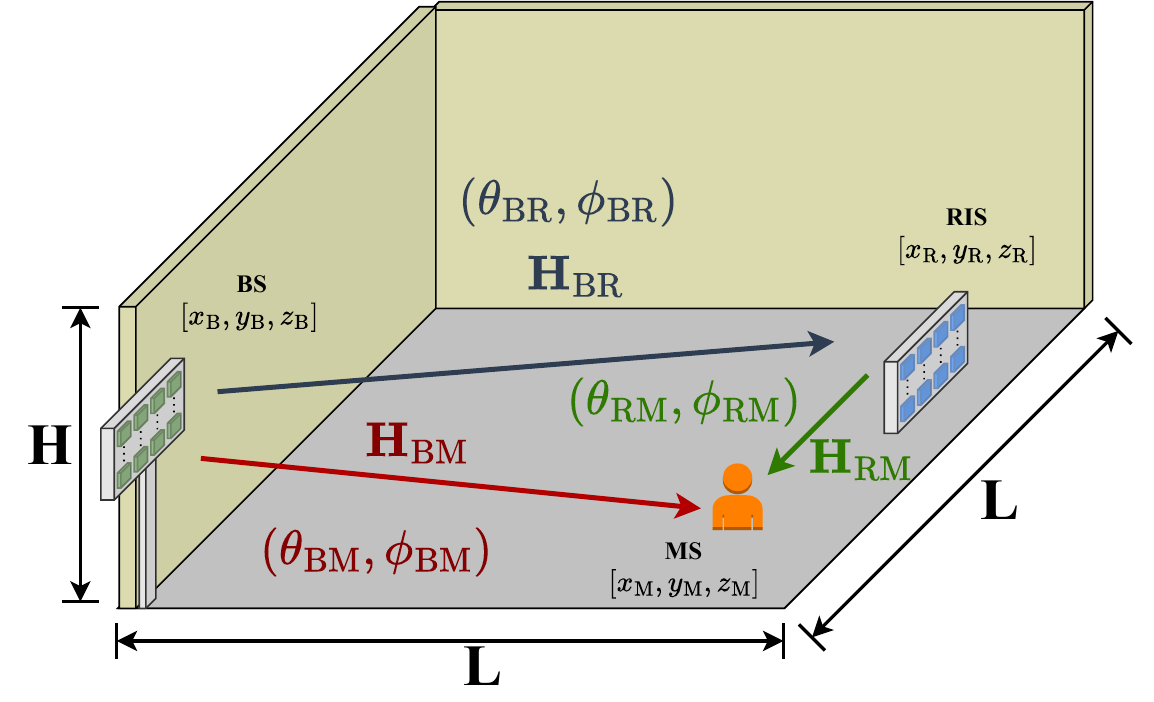}
    \caption{Indoor scenario for RIS-aided mmWave MIMO systems. The environment is modeled as a square with sides of length (L) and height (H).}
    \label{fig:SimulatedScenario}
    \vspace{-5mm}
\end{figure}

We consider a 3D indoor environment, as depicted in Fig.~\ref{fig:SimulatedScenario}, which includes a multi-antenna \ac{BS}, a \ac{RIS}, and a \ac{MS}. These are located at $\hat{\boldsymbol{\rho}}_\textsubscript{B} = [x_\text{B}, y_\text{B}, z_\text{B}]^\text{T}$, $\hat{\boldsymbol{\rho}}_\textsubscript{R} = [x_\text{R}, y_\text{R}, z_\text{R}]^\text{T}$, and $\hat{\boldsymbol{\rho}}_\textsubscript{M} = [x_\text{M}, y_\text{M}, z_\text{M}]^\text{T}$, and rotated $\text{r}_\textsubscript{B} = [\alpha_\textsubscript{B}, \beta_\textsubscript{B}, \gamma_\textsubscript{B}]$, $\text{r}_\textsubscript{R} = [\alpha_\textsubscript{R}, \beta_\textsubscript{R}, \gamma_\textsubscript{R}]$, and $\text{r}_\textsubscript{M}=[\alpha_\textsubscript{M}, \beta_\textsubscript{M}, \gamma_\textsubscript{M}]$, respectively. We assume a \ac{UPA} for the antennas/elements, which are parallel to the $x$-$z$ plane. The \ac{OFDM} downlink received signal model at the \ac{MS}, where $n$ and $t$ denote the \ac{OFDM} subcarrier and symbol index, respectively, is assumed as follows:
\begin{equation}
    y^{u_1,u_2}_{n,t} =  \sum_{v_1,v_2} (H^{\text{BM},v_1,v_2}_{n,u_1,u_2} + H^{\text{BRM},v_1,v_2}_{n,t,u_1,u_2}) x^{v_1,v_2}_{n,t} + w^{u_1,u_2}_{n,t},
    \label{eq:signal_model}
\end{equation}
\noindent{where $\mathbf{H}^{\text{BM}}$ represents channel tensor of the direct communication link between the BS and the MS, $\mathbf{H}^{\text{BRM}}$ is the channel tensor of the link between the \ac{BS}, \ac{RIS} and the \ac{MS}, $\mathbf{x}$ is the transmitted pilot tensor, and $w^{u_1,u_2}_{n,t}$ is \ac{AWGN} following $\mathcal{CN}(0, N_0)$.} Furthermore, the $v_1$ and $v_2$ indices represent the transmit antennas along each dimension (assuming a uniform rectangular array) and go up to $N_{\text{B}_1}-1$ and $N_{\text{B}_2}-1$, respectively. Similarly, the $u_1$ and $u_2$ indices are related to the receive antennas.

The BS-MS channel tensor, denoted as $\mathbf{H}^\text{BM}$, is complex valued of size $N_c \times N_{\text{M}_1} \times N_{\text{M}_2} \times N_{\text{B}_1} \times N_{\text{B}_2}$, where $N_c$ is the number of subcarriers, $N_{\text{B}_1}$ and $N_{\text{B}_2}$ represent the number of antennas along each direction of the rectangular array of the \ac{BS} (similarly defined for $N_{\text{M}_1}$ and $N_{\text{M}_2}$ at the \ac{MS}). Its structure is modeled as a \ac{LOS} channel
\begin{equation}
    H^{\text{BM},v_1,v_2}_{n,u_1,u_2} = b_\text{BM} e^{j n \omega_\text{BM}} e^{j u_1 \psi^\text{BM}_1} e^{j u_2 \psi^\text{BM}_2} e^{j v_1 \varsigma^\text{BM}_1} e^{j v_2 \varsigma^\text{BM}_2}  ,
    \label{eq:h_bm}
\end{equation}
where the $\psi^\text{BM}$ and the $\varsigma^\text{BM}$ are, respectively, the angle of arrival and angle of departure spacial frequencies from the \ac{BS} to \ac{MS} direct link. These are defined in the following fashion
\begin{equation}
    \psi^\text{BM}_1 = \pi \sin{\theta'_\text{MB}}\cos{\phi'_\text{MB}}, \hspace{0.5cm}
    \psi^\text{BM}_2 = \pi\cos{\theta'_\text{MB}},
    \label{eq:psibm}
\end{equation}
\begin{equation}
    \varsigma^\text{BM}_1 = \pi\sin{\theta'_\text{BM}}\cos{\phi'_\text{BM}},
    \hspace{0.5cm}
    \varsigma^\text{BM}_2 = \pi\cos{\theta'_\text{BM}},
    \label{eq:varsbm}
\end{equation}
where $\theta_\text{BM}$ and $\phi_\text{BM}$ represent the zenith and azimuth departure angles from the BS to the MS, respectively. Similarly, $\theta_\text{MB}$ and $\phi_\text{MB}$ represent the zenith and azimuth arrival angles to the \ac{MS} from the \ac{BS}, respectively. Futhermore,
\begin{equation}
    \omega_\text{BM} = -2\pi(\tau_{\text{BM}} + \tau_0)f_\text{sc},
    \label{eq:omega}
\end{equation}
where $\tau_{\text{BM}}$ is the propagation delay between BS and MS, $\tau_0$ is the clock timing offset, and $f_\text{sc}$ is the subcarrier spacing. The propagation path loss, denoted as $b_\text{BM}$, is given by 
\begin{equation}
    b_\text{BM} = \sqrt{\frac{c}{f_\text{c} 4 \pi d_\text{BM}}}e^{j\angle b_\text{BM}},
    \label{eq:pathloss}
\end{equation}
where $d_\text{BM}$ is the distance between BS-MS, $c$ is the speed of light, $f_\text{c}$ is the central frequency, and $\angle b_\text{BM}$ is the path phase contribution.

The BS-RIS-MS channel is derived by cascading the BS-RIS channel $\mathbf{H}^\text{BR}$ and the RIS-MS channel $\mathbf{H}^\text{RM}$. The compound channel is modeled in the following way
\begin{multline}
    H^{\text{BRM},v_1,v_2}_{n,t,u_1,u_2} = b_\text{BR}b_\text{RM} e^{j n \omega_\text{BRM}} e^{j u_1 \psi^\text{RM}_1} e^{j u_2 \psi^\text{RM}_2} \\
                \left( \sum_{k_1,k_2} \Omega_{k_1,k_2,t} e^{j k_1 \vartheta_1} e^{j k_2 \vartheta_2}\right)  e^{j v_1 \varsigma^\text{BR}_1} e^{j v_2 \varsigma^\text{BR}_2},
    \label{eq:BRM_channel}
\end{multline}
where the $b$, $\psi$, and $\varsigma$ parameters are analogous to those in (\ref{eq:h_bm}). Also, $\vartheta_1 = \psi^\text{BR}_1 + \varsigma^\text{RM}_1$, $\vartheta_2 = \psi^\text{BR}_2 + \varsigma^\text{RM}_2$, and $\Omega_{k_1,k_2} = e^{j \omega^\text{RIS}_{k_1,k_2,t}}$ is the \ac{RIS} phase shift at element $(k_1,k_2)$ and symbol $t$. Finally, $\omega_\text{BRM} = -2\pi(\tau_{\text{BR}} + \tau_{\text{RM}} + \tau_0)f_\text{sc}$. The \ac{RIS} is rectangular with dimensions $N_{\text{R}_1} \times N_{\text{R}_2}$, thus the total number of elements is $N_\text{R} = N_{\text{R}_1} N_{\text{R}_2} $.

\section{Fundamental Bounds}
In this paper, we assume that \ac{CE} is applied, allowing us to retrieve near-perfect channel estimates with some estimation error. This could be achieved, for example, by leveraging various mechanisms such as error correction, feedback techniques, and advanced signal processing methods, including Deep Neural Networks and sparse signal recovery \cite{he2022simultaneous,an2024two,bayraktar2024ris,huang2019indoor}. The estimated channel parameters are given by
\begin{equation}
    \hat{\boldsymbol{\eta}} = \boldsymbol{\eta} + \textbf{w},
    \label{eq:ce_error}
\end{equation}
\noindent{where $\textbf{w} \sim \mathcal{N}(0, \sigma^2\textbf{I})$ represents the parameters estimation error, and $\boldsymbol{\eta}$ are the channel parameters of interest, given by}
\begin{multline}
    \boldsymbol{\eta} = [ {\psi}^\text{BM}_1, {\psi}^\text{BM}_2, {\psi}^\text{RM}_1, {\psi}^\text{RM}_2, {\varsigma}^\text{BM}_1, {\varsigma}^\text{BM}_2, \\
    {\varsigma}^\text{RM}_1, {\varsigma}^\text{RM}_2, {\omega}_\text{BM}, {\omega}_\text{BRM}, {b}_\text{BM}, {b}_\text{RM} ].
    \label{eq:channel_param}
\end{multline}

Our proposed framework focuses on estimating the position and rotation angles of the \ac{MS}. We aim to compare our results with the \ac{CRLB} for the \ac{MS}'s position and orientation, referred to as the \ac{PEB} and \ac{REB}, respectively. The \ac{PEB} and \ac{REB} can be derived by differentiating the channel parameters \acs{PDF} with respect to the parameter vector of interest \cite{ghazalian2024joint, liu2022cramer}, given by $\boldsymbol{\zeta} = [ x_\text{M}, y_\text{M}, z_\text{M}, \alpha_\text{M}, \beta_\text{M}, \gamma_\text{M}]$, as follows
\begin{equation} 
    \scalebox{1}{$
    \setlength{\arraycolsep}{2pt}
    \medmuskip = 1mu
    \text{CRLB} = 
    \mathbb{E}\left[- \frac{\partial^2 \ln p(\boldsymbol{\hat{\eta}} ; \boldsymbol{\zeta})}{\partial \boldsymbol{\zeta}}^2 \right]^{-1}= 
    \left[ \frac{1}{\sigma^2} \left\{ \frac{\partial\boldsymbol{\eta}}{\partial \boldsymbol{\zeta}} \frac{\partial\boldsymbol{\eta}^\text{T}}{\partial \boldsymbol{\zeta}} \right\}\right]^{-1},
    \label{eq:CRLB}
    $}
\end{equation} 
\begin{equation}
    \scalebox{1}{$
    \setlength{\arraycolsep}{2pt}
    \medmuskip = 1mu
    \text{PEB} = \sqrt{\text{tr}(\text{CRLB}_{1:3,1:3})}, \hspace{0.2cm}
    \text{REB} = \sqrt{\text{tr}(\text{CRLB}_{4:6,4:6})},
    \label{eq:Bounds}
    $}
\end{equation}
\noindent{the elements to compute \eqref{eq:CRLB}} can be found in the Appendix.

\section{Graphical Models and Bayesian Inference}
Bayesian networks are probabilistic graphical models that represent the joint probability distribution of a set of random variables through their conditional interdependencies. They utilize prior knowledge to infer the likelihood of events and make predictions. These networks are useful in solving localization problems by incorporating information about the system such as the channel model and error distributions.

Depicted as \acp{DAG}, Bayesian networks have vertices representing random variables and edges denoting its interdependencies~\cite{lauritzen1996graphical}. In a \ac{DAG}, each variable is conditionally independent of its non-descendants given its parents. This allows the joint \ac{PDF} to be factorized, simplifying computations. For a \ac{DAG} $\mathcal{D} = (V,E)$, the joint \ac{PDF} of the random variables $X_{v}$, $v \in V$, where $\pa(v)$ denotes the parents of $v$~\cite{koller2009probabilistic}, is given by
\begin{equation}
    \label{eq:pdf_bayes}
    p(V) = \prod_{v \in V} p(v | \pa(v)).
\end{equation}

When dealing with high-dimensional and complex distributions, finding the exact joint distribution that satisfies the model conditions can be difficult. Therefore, approximate inference methods like \ac{MCMC} are often employed. \ac{MCMC} methods, based on Bayes's theorem, estimate the posterior distribution of model parameters to support the analysis and prediction~\cite{metropolis1949monte}. These methods generate sequences of random samples to approximate target distributions. The \ac{NUTS} is an efficient \ac{MCMC} method that avoids re-exploration and reduces simulation time by using a recursive algorithm to identify candidate points in the target distribution~\cite{hoffman2014no}.

\subsection{Localization Graphical Model}
\begin{figure}[!t]
    \centerline{\includegraphics[width=\columnwidth]{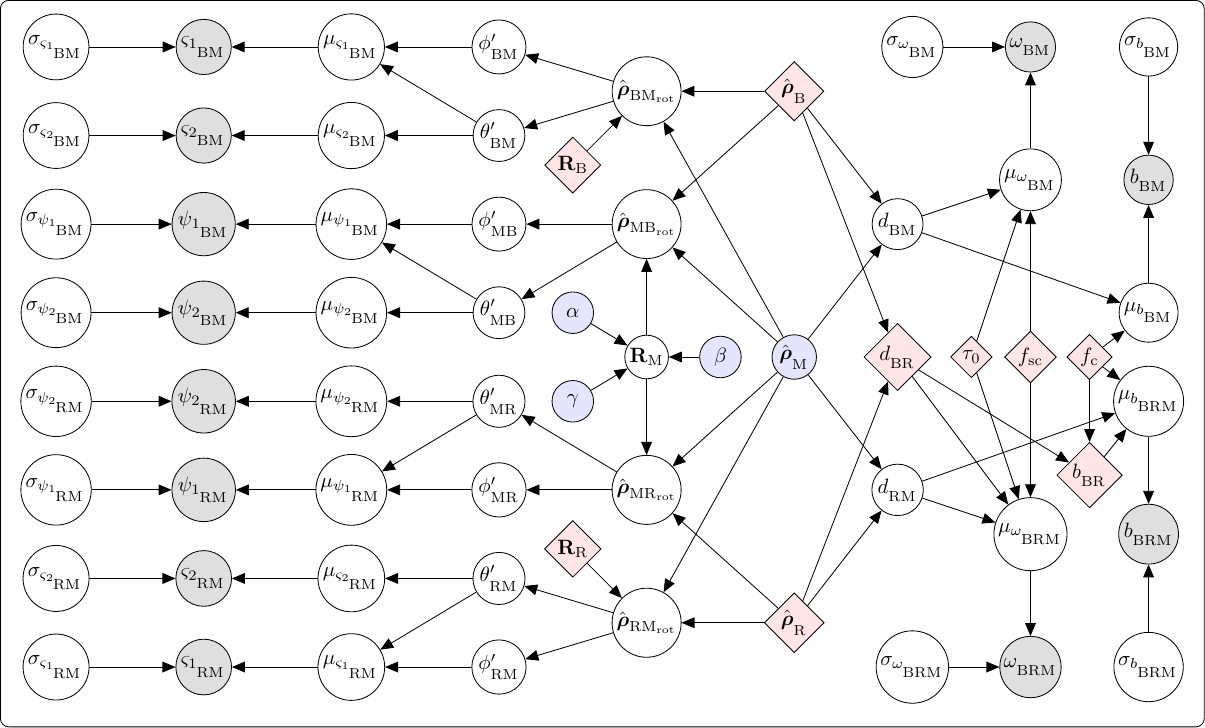}}
    \caption{Graphical model of the source localization.}
    \label{fig:GraphicalModel}
    \vspace{-5mm}
\end{figure}

The directed acyclic graph model, shown in Fig.~\ref{fig:GraphicalModel}, represents our system. The parameters within the circles correspond to random variables whose distributions are derived from prior knowledge. The grey circles correspond to our observed data, and the blue color highlights the random variables of interest, which are the coordinates of the \ac{MS} and its rotation given by $\alpha$, $\beta$ and $\gamma$. On the other hand, the red diamonds represent the known parameters, namely $\hat{\boldsymbol{\rho}}_\textsubscript{B}$, $\hat{\boldsymbol{\rho}}_\textsubscript{R}$, $d_\text{BR}$, $\mathbf{R}_\text{B}$, $\mathbf{R}_\text{R}$, $f_c$, $f_\text{sc}$ and $\tau_0$. These represent the coordinates of the \ac{BS}, the coordinates of the \ac{RIS}, the distance between the \ac{BS} and \ac{RIS}, the rotation matrix for the \ac{BS}, the rotation matrix for the \ac{RIS}, the central frequency, the subcarrier spacing, and the clock offset, respectively. These parameters are considered known by the system for all simulations. 

The assumptions on the random variables' distribution constituting the respective Bayesian network are given in Table~\ref{table:rv_bayesian}, where $\sim$ indicates that a random variable follows a specific distribution. We use a uniform distribution to initialize our prior knowledge of $(x_\text{M}, y_\text{M}, z_\text{M}, \alpha_\text{M}, \beta_\text{M}, \gamma_\text{M})$. This approach ensures that the sample space covers all the deployment scenario and possible rotations, once the target can be at any place of the environment with any rotation with same probability. The upper bound $L$ is the side length of our deployment scenario, $H$ is the height of the environment, and $\epsilon$ represents the maximum possible rotation angle. The covariance matrix for the observed data is given by $\Sigma_{\text{data}}$, where its main diagonal comprises a vector given by $\boldsymbol{\sigma}_{\eta}^2$, representing the variance of each observed data in \eqref{eq:channel_param}. We model this variance with a small shape and scale in the inverse gamma distribution. This choice reduces the step size in the sampling process, simplifying the examination around the average.
\begin{table}[!t]
    \caption{Random variables of the Bayesian network\tablefootnote{$\mathcal{U}(l,u)$ represents a uniform distribution, where $l$ is the lower bound and $u$ is the upper bound, \text{InverseGamma}($\alpha, \beta$) represents a inverse gamma distribution, where $\alpha$ is the shape parameter and $\beta$ the scale parameter, and $ij$ are indices from a group $\mathcal{G} = \{(i, j) \in \{\text{BM, MB, RM, MR}\} \}$.}}
    \begin{center}
        \begin{tabular}{c l}
            \hline
            \textbf{Variable} & \textbf{Description} \rule[-1ex]{0pt}{4ex}\\
            \hline
            $x_\text{M}, y_\text{M}$ & $\sim$ $\mathcal{U}(0, L)$ \rule[0ex]{0pt}{2.5ex} \\
            $z_\text{M}$ & $\sim$ $\mathcal{U}(0, H)$ \rule[0ex]{0pt}{2.5ex} \\
            $\alpha_\text{M}, \beta_\text{M}, \gamma_\text{M}$ & $\sim$ $\mathcal{U}(0, \epsilon)$ \rule[0ex]{0pt}{2.5ex} \\
            $\mathbf{R}_\text{M}$ & $\sim$ Equation \eqref{eq:rot_matrix} \rule[0ex]{0pt}{2.5ex} \\
            $d_\text{BM}$ & $\sim$ $ ||\hat{\boldsymbol{\rho}}_{\text{B}} - \hat{\boldsymbol{\rho}}_{\text{M}}|| $ \rule[0ex]{0pt}{2.5ex} \\
            $d_\text{RM}$ & $\sim$ $ ||\hat{\boldsymbol{\rho}}_{\text{R}} - \hat{\boldsymbol{\rho}}_{\text{M}}|| $ \rule[0ex]{0pt}{2.5ex} \\
            $\hat{\boldsymbol{\rho}}_{{ij}_{\text{rot}}}$ & $\sim$ $ \mathbf{R}_{j}^{T} 
            (\hat{\boldsymbol{\rho}}_{j} - \hat{\boldsymbol{\rho}}_{i})$ \rule[0ex]{0pt}{2.5ex} \\
            $\theta'_{ij}$ & $\sim$ $\arccos(\hat{\boldsymbol{\rho}}_{{ij}_{\text{rot}}}[2])$ \rule[0ex]{0pt}{2.5ex} \\
            $\phi'_{ij}$ & $\sim$ $\arctan2 \left( \hat{\boldsymbol{\rho}}_{{ij}_{\text{rot}}}[1], \hat{\boldsymbol{\rho}}_{{ij}_{\text{rot}}}[0] \right)$ \rule[0ex]{0pt}{2.5ex} \\
            $\mu_{\psi_{1_{\text{\{BM,RM\}}}}}$ & $\sim$ $ -\pi \sin(\theta'_{\text{\{MB,MR\}}}) \cos(\phi'_{\text{\{MB,MR\}}}) $ \rule[0ex]{0pt}{2.5ex} \\
            $\mu_{\psi_{2_{\text{\{BM,RM\}}}}}$ & $\sim$ $ -\pi \cos(\theta'_{\text{\{MB,MR\}}}) $ \rule[0ex]{0pt}{2.5ex} \\
            $\mu_{\varsigma_{1_{\text{\{BM,RM\}}}}}$ & $\sim$ $ -\pi \sin(\theta'_{\text{\{BM,RM\}}}) \cos(\phi'_{\text{\{BM,RM\}}}) $ \rule[0ex]{0pt}{2.5ex} \\
            $\mu_{\varsigma_{2_{\text{\{BM,RM\}}}}}$ & $\sim$ $ -\pi \cos(\theta'_{\text{\{BM,RM\}}}) $ \rule[0ex]{0pt}{2.5ex} \\
            $\mu_{\omega_\text{BM}}$ & $\sim$ $-2 \pi ((d_\text{BM}/c) + \tau_0)f_\text{sc}$ \rule[0ex]{0pt}{2.5ex} \\
            $\mu_{\omega_\text{BRM}}$ & $\sim$ $-2 \pi (((d_\text{BR} + d_\text{RM})/c) + \tau_0)f_\text{sc}$ \rule[0ex]{0pt}{2.5ex} \\
            $\mu_{b_\text{BM}}$ & $\sim$ $(c / f4\pi d_\text{BM})$ \rule[0ex]{0pt}{2.5ex} \\
            $\mu_{b_\text{BRM}}$ & $\sim$ $b_\text{BR} (c / f4\pi d_\text{RM})$ \rule[0ex]{0pt}{2.5ex} \\  $\sigma_{\boldsymbol{\eta}}^2$ & $\sim$ $\text{InverseGamma}(0.001,0.001)$ \rule[0ex]{0pt}{2.5ex} \\  
            \hline
        \end{tabular}
        \label{table:rv_bayesian}
        \vspace{-7mm}
    \end{center}
\end{table}
\section{Simulation Results}
In our simulations, we implement the indoor deployment scenario introduced in Section~\ref{sec:system}. The scenario is a square area with a side length of \(15\) meters and \(5\) meters height, containing  a \ac{RIS} positioned at $\hat{\boldsymbol{\rho}}_\textsubscript{R} = [7.5, 15, 4]^\text{T}$ and a base station at $\hat{\boldsymbol{\rho}}_\textsubscript{R} = [0, 0, 5]^\text{T}$, both with known orientations. The channels are generated based on \eqref{eq:signal_model}, where \ac{CE} is applied to obtain near-perfect channel estimates, incorporating some estimation error as defined by \eqref{eq:ce_error}. For efficiency, the orientation of the \ac{MS} is constrained between $0$ and $ \pi/4$; orientations beyond this range would require a denser point sampling and thus increase simulation time. The localization estimator is implemented using the PyMC package~\cite{pymc2023}, employing the \ac{NUTS} algorithm to sample from the posterior distribution of coordinates and rotations. 
We feed \ac{NUTS} with $50$ data samples drawn from a Gaussian distribution, $\mathcal{N}(\boldsymbol{\hat{\eta}}, \sigma_{\text{sp}}^2 \mathbf{I})$, based on the estimated parameters defined in \eqref{eq:ce_error} and $\sigma_{\text{sp}}^2 = 10^{-3}$ to reflect confidence on the \ac{CE} accuracy. In each simulation run, \ac{NUTS} operates with $4$ chains, drawing $2500$ samples per chain. The step size is tuned to $1500$ to achieve an approximate sample acceptance rate of $90\%$. Targets are placed randomly within the simulated area, each assigned a random position and orientation. 

\begin{figure}[!t]
    \centerline{\includegraphics[width=0.95\columnwidth]{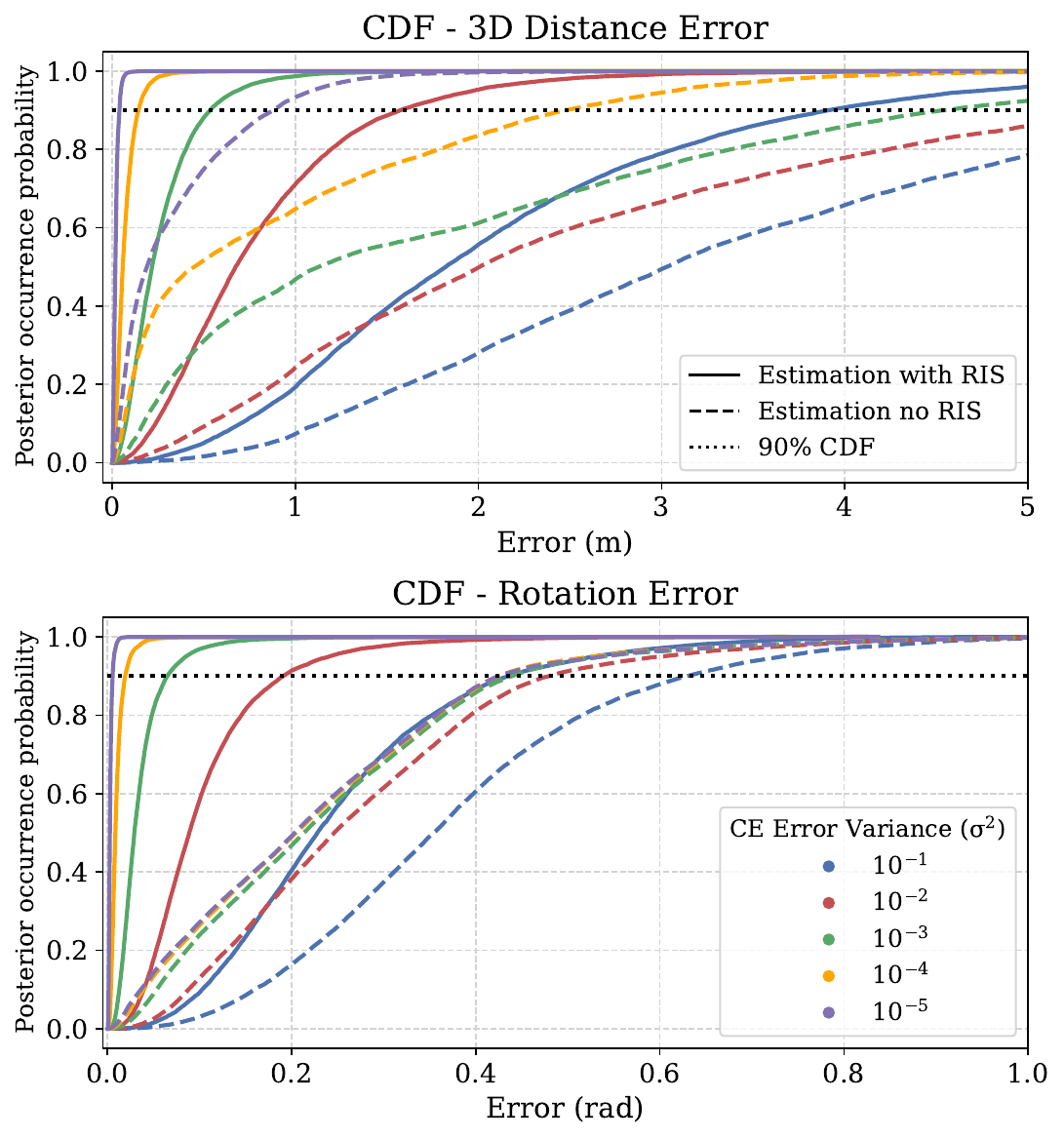}}
    \caption{CDF of the 6D error for the proposed source localization.}
    \label{fig:CDF}
    \vspace{-3mm}
\end{figure}

The \ac{CDF} for various values of \ac{CE} error is illustrated in Fig.~\ref{fig:CDF}. The results are averaged over $30.000$ simulation runs. In each iteration, the target assumes a different position and orientation, and the algorithm is fed with new samples. The black dotted line indicates the $90$th percentile of the \ac{CDF}, with its value is presented in Table~\ref{tab:cdf}. This threshold indicates that $90$\% of the estimates are achieved with an error less than or equal to this value.
\begin{table}[!t]
    \addtolength{\tabcolsep}{-0.2em}
    \centering
    \caption{90\% marker of the CDF varying the \ac{CE} error variance.}
    \label{tab:cdf}
    \begin{tabular}{l l c c c c c}
        \cmidrule[1pt](lr){1-7}
        \multirow{2}{*}{\textbf{Metric}} & \multirow{2}{*}{\textbf{Scenario}} & \multicolumn{5}{c}{\boldmath$\sigma^2$} \\
        \cmidrule[1pt](lr){3-7}
        & & $10^{-1}$ & $10^{-2}$ & $10^{-3}$ & $10^{-4}$ & $10^{-5}$ \\
        \cmidrule[1pt](lr){1-7}
        \multirow{2}{*}{\textbf{PEB} (m)} & RIS & 3.91063 & 1.57071 & 0.53481 & 0.14384 & 0.03853 \\
        & No RIS & 6.44283 & 5.55535 & 4.53914 & 2.47313 & 0.87843  \\
        \cmidrule[1pt](lr){1-7}
        \multirow{2}{*}{\textbf{REB} (rad)} & RIS & 0.43547 & 0.19091 & 0.06510 & 0.01920 & 0.00547 \\
        & No RIS & 0.63008 & 0.47993 & 0.43972 & 0.42581 & 0.42543  \\
        \cmidrule[1pt](lr){1-7}
    \end{tabular}
    \vspace{-6mm}
\end{table}
\begin{figure*}[!t]
    \centerline{\includegraphics[width=2\columnwidth]{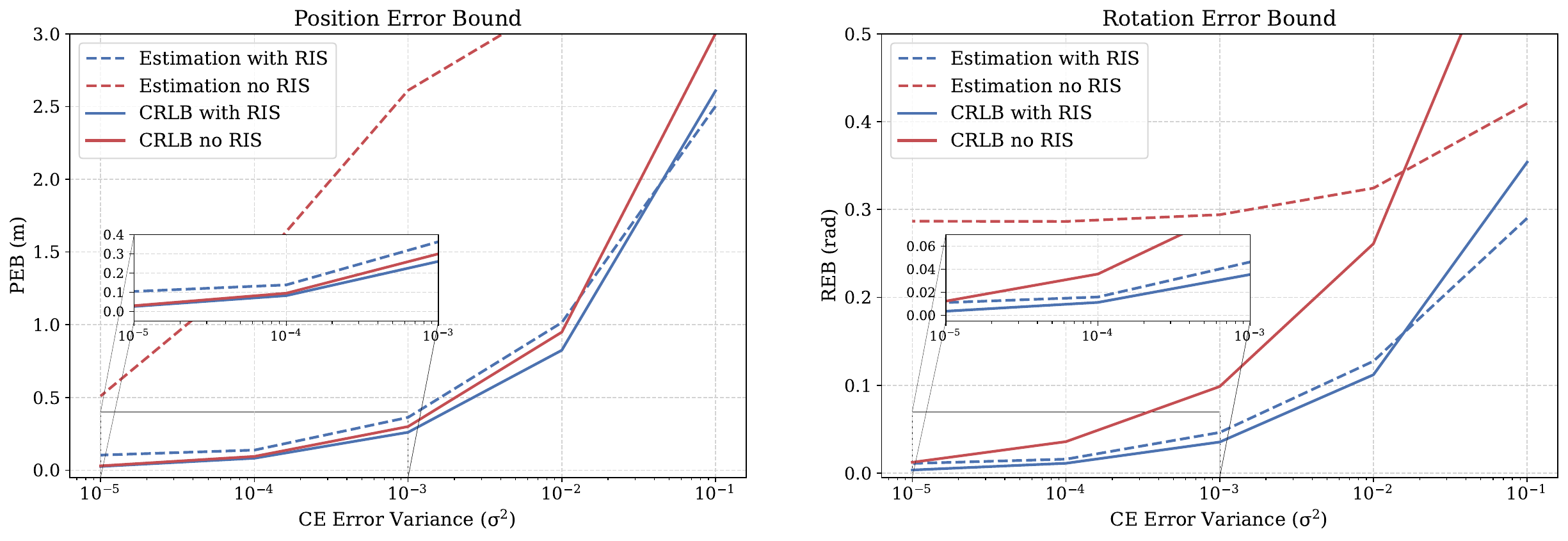}}
    \caption{PEB and REB for the proposed scenario varying the \ac{CE} error variance.}
    \label{fig:PEB_REB}
    \vspace{-5mm}
\end{figure*}
The results show that scenarios without \ac{RIS} shows lower accuracy in both position and rotation, compared to scenarios with \ac{RIS}. The additional channel path provides more information about the target than simply relying on a \ac{LOS} path. We observe a substantial improvement in rotation accuracy. Decreasing the error associated with the \ac{CE} considerably increases the accuracy of the estimation of position and rotation in all scenarios. In cases with a lower \ac{CE} error, such as $\sigma^2 = 10^{-3}$, and the presence of \ac{RIS}, we observe an average error of $0.53$~m in position and $0.06$~rad in rotation, compared to $4.54$~m and $0.44$~rad without \ac{RIS}.

Using the results from previous simulations, we calculate the \ac{PEB} and \ac{REB} in Fig.~\ref{fig:PEB_REB} and compare them with the theoretical bounds from \eqref{eq:Bounds}. The results from scenarios with \ac{RIS} are closely aligned with the bounds. Specifically, at low $\sigma^2$ values (e.g., below $10^{-2}$), the \ac{PEB} and \ac{REB} in \ac{RIS}-enabled scenarios are closely matching from the bounds. However, the same behavior does not apply in scenarios without \ac{RIS}, where performance deviates from the bound. This difference occurs because, without the additional path provided by \ac{RIS}, the algorithm lacks sufficient information to effectively estimate all the variables in the system. This limitation is particularly evident in rotation estimation, where performance stabilizes around $0.3$~rad, as the algorithm relies on guessing it's rotation. 

The presence of \ac{RIS} significantly enhances parameter estimation by providing more information about the \ac{MS} position and rotation. As shown in Fig.~\ref{fig:PEB_REB}, our framework achieves near-perfect estimations, with position errors below $0.4$~m and rotation errors below $0.05$~rad when errors are below $\sigma^2~=~10^{-2}$. In high \ac{CE} error scenarios, the performance of both scenarios is considerably degraded. Even with \ac{RIS}, the high parameter error prevents the algorithm from reliably estimating position and rotation, leading it to rely on approximate guesses for rotation. This limitation may be mitigated by increasing the number of samples for the algorithm or the number of draws for \ac{NUTS}. However, the approach would also increase processing time or complexity.

\section{Conclusions}
We have developed a 6D Bayesian-based localization framework for estimating both position and rotation in an indoor \ac{RIS}-aided environment. By utilizing a probabilistic graphical model and the \ac{NUTS}, our approach accurately approximates the posterior distribution of localization parameters, using estimated channel parameters for improved precision. Through the derivation of the \ac{CRLB}, we have quantified the system's accuracy with \ac{PEB} and \ac{REB}, setting it as a benchmark for evaluation. Our comparative analysis with scenarios without \ac{RIS} highlights the performance improvements enabled by \ac{RIS}, particularly in enhancing the accuracy of \ac{MS} rotation estimation. This work demonstrates the potential of probabilistic graphical models and Bayesian methods for indoor localization problems, providing a robust framework that operates without the need for training, iterative learning, or multiple passes, achieving accurate estimates in a single snapshot. Future research directions include applying different \ac{CE} algorithms to analyze their impact, exploring dynamic scenarios (e.g. iterative estimations for moving targets), examining the impact of multi-path components, and analyzing other data fusion techniques, such as visual data, to enhance and refine the proposed method. 

\section*{Acknowledgment}
The work was supported by the Research Council of Finland (former Academy of Finland) 6G Flagship Program (Grant Number: 346208). We also thank Prof. Mikko Sillanpää, Enrique Pinto, Hamza Djelouat, and Reijo Leinonen for the productive discussions.

\appendices
\section{\ac{CRLB} derivation}
To derive the \ac{CRLB} in \eqref{eq:CRLB}, we need to compute $\frac{\partial\boldsymbol{\eta}}{\partial \boldsymbol{\zeta}}$. By using \eqref{eq:psibm}, \eqref{eq:varsbm}, \eqref{eq:omega} and \eqref{eq:pathloss}, along with the parameter relationships given by \eqref{eq:rotation_rz_ry}, \eqref{eq:rotation_rx} and \eqref{eq:theta_phi'}, the elements of the \ac{CRLB} can be computed. Before deriving the parameters, we first introduce the following auxiliary variables
\begin{equation}
    \scalebox{0.95}{$
    \setlength{\arraycolsep}{2pt}
    \medmuskip = 1mu
    \boldsymbol{\chi}_\text{BM} \triangleq \hat{\boldsymbol{\rho}}_\textsubscript{M} - \hat{\boldsymbol{\rho}}_\textsubscript{B}, \hspace{0.1cm}
    \boldsymbol{\chi}_\text{RM} \triangleq \hat{\boldsymbol{\rho}}_\textsubscript{M} - \hat{\boldsymbol{\rho}}_\textsubscript{R}, \hspace{0.1cm} d_\text{BM} \triangleq ||\hat{\boldsymbol{\rho}}_\textsubscript{M} - \hat{\boldsymbol{\rho}}_\textsubscript{B}||
    $}
\end{equation}
\begin{equation}
    \scalebox{0.95}{$
    \setlength{\arraycolsep}{2pt}
    \medmuskip = 1mu
    d_\text{RM} \triangleq ||\hat{\boldsymbol{\rho}}_\textsubscript{M} - \hat{\boldsymbol{\rho}}_\textsubscript{R}||, \hspace{0.1cm} \textbf{K}_{\text{MB}} = \mathbf{R}_\text{M} \frac{-\boldsymbol{\chi}_\text{BM}}{d_\text{BM}}, \hspace{0.1cm} 
    \textbf{K}_{\text{MR}} = \mathbf{R}_\text{M} \frac{-\boldsymbol{\chi}_\text{RM}}{d_\text{RM}},
    $}
\end{equation}
\begin{equation}
    \scalebox{1}{$
    \setlength{\arraycolsep}{2pt}
    \medmuskip = 1mu
    \textbf{K}_{\text{BM}} = \mathbf{R}_\text{B} \frac{\boldsymbol{\chi}_\text{BM}}{d_\text{BM}} \hspace{0.3cm} \text{and}\hspace{0.3cm}
    \textbf{K}_{\text{RM}} = \mathbf{R}_\text{R} \frac{\boldsymbol{\chi}_\text{RM}}{d_\text{RM}},
    $}
\end{equation}
as well $\mathbf{u}_1 \triangleq [1, 0, 0]^\text{T}$, $\mathbf{u}_2 \triangleq [0, 1, 0]^\text{T}$ and $\mathbf{u}_3 \triangleq [0, 0, 1]^\text{T}$, to rewrite \eqref{eq:theta_phi'} as
\begin{equation}
    \scalebox{0.9}{$
    \setlength{\arraycolsep}{2pt}
    \medmuskip = 1mu
    \theta'_\text{MB} = \arccos \left( \mathbf{u}_3^\text{T} \textbf{K}_{\text{MB}}\right), \phi'_\text{MB} = \arctan2 \left( \mathbf{u}_2^\text{T} \textbf{K}_{\text{MB}}, \mathbf{u}_1^\text{T} \textbf{K}_{\text{MB}}\right),
    $}
\end{equation}
\begin{equation}
    \scalebox{0.9}{$
    \setlength{\arraycolsep}{2pt}
    \medmuskip = 1mu
    \theta'_\text{MR} = \arccos \left( \mathbf{u}_3^\text{T} \textbf{K}_{\text{MR}}\right), \phi'_\text{MR} = \arctan2 \left( \mathbf{u}_2^\text{T} \textbf{K}_{\text{MR}}, \mathbf{u}_1^\text{T} \textbf{K}_{\text{MR}}\right),
    $}
\end{equation}
\begin{equation}
    \scalebox{0.9}{$
    \setlength{\arraycolsep}{2pt}
    \medmuskip = 1mu
    \theta'_\text{BM} = \arccos \left( \mathbf{u}_3^\text{T} \textbf{K}_{\text{BM}}\right), \phi'_\text{BM} = \arctan2 \left( \mathbf{u}_2^\text{T} \textbf{K}_{\text{BM}}, \mathbf{u}_1^\text{T} \textbf{K}_{\text{BM}}\right),
    $}
\end{equation}
\begin{equation}
    \scalebox{0.9}{$
    \setlength{\arraycolsep}{2pt}
    \medmuskip = 1mu
    \theta'_\text{RM} = \arccos \left( \mathbf{u}_3^\text{T} \textbf{K}_{\text{RM}}\right), \phi'_\text{RM} = \arctan2 \left( \mathbf{u}_2^\text{T} \textbf{K}_{\text{RM}}, \mathbf{u}_1^\text{T} \textbf{K}_{\text{RM}}\right).
    $}
\end{equation}

Then we have the following derivatives:
\begin{equation}
    \setlength{\arraycolsep}{2pt}
    \medmuskip = 1mu
    \frac{\partial\theta'_\text{MB}}{\partial \hat{\boldsymbol{\rho}}_\text{M}} = \frac{-\mathbf{u}_3^\text{T} \mathbf{R}_\text{M} (-\textbf{A}_\text{BM})}{\sqrt{1 - (\mathbf{u}_3^\text{T}\textbf{K}_{_\text{MB}})^2}}, 
    \frac{\partial\theta'_\text{MR}}{\partial \hat{\boldsymbol{\rho}}_\text{M}} = \frac{-\mathbf{u}_3^\text{T} \mathbf{R}_\text{M} (-\textbf{A}_\text{RM})}{\sqrt{1 - (\mathbf{u}_3^\text{T}\textbf{K}_{_\text{MR}})^2}},
\end{equation}
\begin{equation}
    \setlength{\arraycolsep}{2pt}
    \medmuskip = 1mu
    \frac{\partial\theta'_\text{BM}}{\partial \hat{\boldsymbol{\rho}}_\text{M}} = \frac{-\mathbf{u}_3^\text{T} \mathbf{R}_\text{B} \textbf{A}_\text{BM}}{\sqrt{1 - (\mathbf{u}_3^\text{T}\textbf{K}_{_\text{BM}})^2}},
    \frac{\partial\theta'_\text{RM}}{\partial \hat{\boldsymbol{\rho}}_\text{M}} = \frac{-\mathbf{u}_3^\text{T} \mathbf{R}_\text{R} \textbf{A}_\text{RM}}{\sqrt{1 - (\mathbf{u}_3^\text{T}\textbf{K}_{_\text{RM}})^2}},
\end{equation}

\begin{multline}
    \setlength{\arraycolsep}{2pt}
    \medmuskip = 1mu
    \frac{\partial\phi'_\text{\{MB,MR\}}}{\partial \hat{\boldsymbol{\rho}}_\text{M}} = \frac{\mathbf{u}_1^\text{T}\textbf{K}_{\text{\{MB,MR\}}}  (\mathbf{R}_\text{M}^\text{T} \mathbf{u}_2)^\text{T}  (-\textbf{A}_\text{\{BM,RM\}})}{\mathbf{u}_1^\text{T}\textbf{K}_{\text{\{MB,MR\}}}^2+\mathbf{u}_2^\text{T}\textbf{K}_{\text{\{MB,MR\}}}^2} - \\ \frac{\mathbf{u}_2^\text{T}\textbf{K}_{\text{\{MB,MR\}}}(\mathbf{R}_\text{M}^\text{T} \mathbf{u}_1)^\text{T}  (-\textbf{A}_\text{\{BM,RM\}})}{\mathbf{u}_1^\text{T}\textbf{K}_{\text{\{MB,MR\}}}^2+\mathbf{u}_2^\text{T}\textbf{K}_{\text{\{MB,MR\}}}^2},
\end{multline}
\begin{multline}
    \setlength{\arraycolsep}{2pt}
    \medmuskip = 1mu
    \frac{\partial\phi'_\text{\{BM,RM\}}}{\partial \hat{\boldsymbol{\rho}}_\text{M}} = \frac{\mathbf{u}_1^\text{T}\textbf{K}_{\text{\{BM,RM\}}}  (\mathbf{R}_\text{\{B,R\}}^\text{T} \mathbf{u}_2)^\text{T}  \textbf{A}_\text{\{BM,RM\}}}{\mathbf{u}_1^\text{T}\textbf{K}_{\text{\{BM,RM\}}}^2+\mathbf{u}_2^\text{T}\textbf{K}_{\text{\{BM,RM\}}}^2} - \\ \frac{\mathbf{u}_2^\text{T}\textbf{K}_{\text{\{BM,RM\}}}(\mathbf{R}_\text{\{B,R\}}^\text{T} \mathbf{u}_1)^\text{T}  \textbf{A}_\text{\{BM,RM\}}}{\mathbf{u}_1^\text{T}\textbf{K}_{\text{\{BM,RM\}}}^2+\mathbf{u}_2^\text{T}\textbf{K}_{\text{\{BM,RM\}}}^2},
\end{multline}
\begin{equation}
    \scalebox{1.1}{$
    \setlength{\arraycolsep}{2pt}
    \medmuskip = 1mu
    \frac{\partial\theta'_\text{MB}}{\partial \textbf{r}_\text{M}} = \frac{-\mathbf{u}_3^\text{T} \frac{\partial \mathbf{R}_\text{M}}{\partial \textbf{r}_\text{M}} \left (\frac{-\boldsymbol{\chi}_\text{BM}}{d_\text{BM}}\right)}{\sqrt{1 - (\mathbf{u}_3^\text{T}\textbf{K}_{_\text{MB}})^2}}, 
    \frac{\partial\theta'_\text{MR}}{\partial \textbf{r}_\text{M}} = \frac{-\mathbf{u}_3^\text{T} \frac{\partial \mathbf{R}_\text{M}}{\partial \textbf{r}_\text{M}} \left (\frac{-\boldsymbol{\chi}_\text{RM}}{d_\text{RM}}\right)}{\sqrt{1 - (\mathbf{u}_3^\text{T}\textbf{K}_{_\text{MR}})^2}},
    $}
\end{equation}
\begin{multline}
    \setlength{\arraycolsep}{2pt}
    \medmuskip = 1mu
    \frac{\partial\phi'_\text{\{MB,MR\}}}{\partial \textbf{r}_\text{M}} = \frac{\mathbf{u}_1^\text{T}\textbf{K}_{\text{\{MB,MR\}}} \mathbf{u}_2^\text{T} \frac{\partial \mathbf{R}_\text{M}}{\partial \textbf{r}_\text{M}} \left (\frac{-\boldsymbol{\chi}_\text{\{MB,MR\}}}{d_\text{\{MB,MR\}}}\right )}{\mathbf{u}_1^\text{T}\textbf{K}_{\text{\{MB,MR\}}}^2+\mathbf{u}_2^\text{T}\textbf{K}_{\text{\{MB,MR\}}}^2} - \\ \frac{\mathbf{u}_2^\text{T}\textbf{K}_{\text{\{MB,MR\}}}\mathbf{u}_1^\text{T} \frac{\partial \mathbf{R}_\text{M}}{\partial \textbf{r}_\text{M}} \left (\frac{-\boldsymbol{\chi}_\text{\{MB,MR\}}}{d_\text{\{MB,MR\}}}\right )}{\mathbf{u}_1^\text{T}\textbf{K}_{\text{\{MB,MR\}}}^2+\mathbf{u}_2^\text{T}\textbf{K}_{\text{\{MB,MR\}}}^2},
\end{multline}
\begin{equation}
    \setlength{\arraycolsep}{2pt}
    \medmuskip = 1mu
    \frac{\partial\theta'_\text{BM}}{\partial \textbf{r}_\text{M}} = \frac{\partial\theta'_\text{RM}}{\partial \textbf{r}_\text{M}} = 
    \frac{\partial\phi'_\text{BM}}{\partial \textbf{r}_\text{M}} = \frac{\partial\phi'_\text{RM}}{\partial \textbf{r}_\text{M}} = 0,
\end{equation}
where
\begin{equation}
    \setlength{\arraycolsep}{2pt}
    \medmuskip = 1mu
     \textbf{A}_\text{\{BM,RM\}} = \frac{\mathbf{I}_3}{d_\text{\{BM,RM\}}} - \frac{\boldsymbol{\chi}_\text{\{BM,RM\}}^\text{T} \otimes \boldsymbol{\chi}_\text{\{BM,RM\}}}{d_\text{\{BM,RM\}}^{3/2}} ,
\end{equation}
\begin{equation}
    \scalebox{1.1}{$
    \setlength{\arraycolsep}{2pt}
    \medmuskip = 1mu
    \frac{\partial \mathbf{R}_\text{M}}{\partial \alpha_\text{M}} = \frac{\partial \mathbf{R}_\text{z}(\alpha_\text{M})}{\partial \alpha_\text{M}}\mathbf{R}_\text{y}(\beta_\text{M})\mathbf{R}_\text{x}(\gamma_\text{M}),
    $}
\end{equation}
\begin{equation}
    \scalebox{1.1}{$
    \setlength{\arraycolsep}{2pt}
    \medmuskip = 1mu
    \frac{\partial \mathbf{R}_\text{M}}{\partial \beta_\text{M}} = \mathbf{R}_\text{z}(\alpha_\text{M})\frac{\partial \mathbf{R}_\text{y}(\beta_\text{M})}{\partial \beta_\text{M}}\mathbf{R}_\text{x}(\gamma_\text{M}),
    $}
\end{equation}
\begin{equation}
    \scalebox{1.1}{$
    \setlength{\arraycolsep}{2pt}
    \medmuskip = 1mu
    \frac{\partial \mathbf{R}_\text{M}}{\partial \gamma_\text{M}} = \mathbf{R}_\text{z}(\alpha_\text{M})\mathbf{R}_\text{y}(\beta_\text{M})\frac{\partial \mathbf{R}_\text{x}(\gamma_\text{M})}{\partial \gamma_\text{M}},
    $}
\end{equation}
\begin{equation}
    \scalebox{0.9}{$
    \setlength{\arraycolsep}{2pt}
    \medmuskip = 1mu
    \frac{\partial \mathbf{R}_\text{z}(\alpha_\text{M})}{\partial \alpha_\text{M}} =
        \begin{bmatrix}
            -\sin(\alpha) & -\cos(\alpha) & 0 \\
            \cos(\alpha) & -\sin(\alpha)  & 0 \\
            0 & 0 & 0
        \end{bmatrix},
    $}
\end{equation}
\begin{equation}
    \scalebox{0.9}{$
    \setlength{\arraycolsep}{2pt}
    \medmuskip = 1mu
    \frac{\partial \mathbf{R}_\text{y}(\beta_\text{M})}{\partial \beta_\text{M}} =
        \begin{bmatrix}
        -\sin(\beta) & 0 & \cos(\beta) \\
            0 & 0  & 0 \\
            -\cos(\beta) & 0 & -\sin(\beta)
        \end{bmatrix},
    $}
\end{equation}
\begin{equation}
    \scalebox{0.9}{$
    \setlength{\arraycolsep}{2pt}
    \medmuskip = 1mu
    \frac{\partial \mathbf{R}_\text{x}(\gamma_\text{M})}{\partial \gamma_\text{M}} =
        \begin{bmatrix}
            0 & 0 & 0 \\
            0 & -\sin(\gamma)  & -\cos(\gamma) \\
            0 & \cos(\gamma) & -\sin(\gamma)
        \end{bmatrix}.
    $}
\end{equation}
The latter expressions are used to compute the following elements:
\begin{equation}
    \setlength{\arraycolsep}{2pt}
    \medmuskip = 1mu
    \frac{\partial \psi^{^\text{\{BM,RM\}}}_1}{\partial \hat{\boldsymbol{\rho}}_\text{M}} = \frac{\partial\psi^{^\text{\{BM,RM\}}}_1}{\partial \theta'_{_\text{\{MB,MR\}}}} \frac{\partial\theta'_{_\text{\{MB,MR\}}}}{\partial \hat{\boldsymbol{\rho}}_\text{M}} - \frac{\partial\psi^{^\text{\{BM,RM\}}}_1}{\partial \phi'_{_\text{\{MB,MR\}}}} \frac{\partial\phi'_{_\text{\{MB,MR\}}}}{\partial \hat{\boldsymbol{\rho}}_\text{M}},
\end{equation}
\begin{equation}
    \setlength{\arraycolsep}{2pt}
    \medmuskip = 1mu
    \frac{\partial \psi^{^\text{\{BM,RM\}}}_2}{\partial \hat{\boldsymbol{\rho}}_\text{M}} = \frac{\partial \psi^{^\text{\{BM,RM\}}}_2}{\partial \theta'_{_\text{\{MB,MR\}}}}\frac{\partial\theta'_{_\text{\{MB,MR\}}}}{\partial \hat{\boldsymbol{\rho}}_\text{M}},
\end{equation}
\begin{equation}
    \setlength{\arraycolsep}{2pt}
    \medmuskip = 1mu
    \frac{\partial \varsigma^{^\text{\{BM,RM\}}}_1}{\partial \hat{\boldsymbol{\rho}}_\text{M}} = \frac{\partial\varsigma^{^\text{\{BM,RM\}}}_1}{\partial \theta'_{_\text{\{BM,RM\}}}} \frac{\partial\theta'_{_\text{\{BM,RM\}}}}{\partial \hat{\boldsymbol{\rho}}_\text{M}} - \frac{\partial\varsigma^{^\text{\{BM,RM\}}}_1}{\partial \phi'_{_\text{\{BM,RM\}}}} \frac{\partial\phi'_{_\text{\{BM,RM\}}}}{\partial \hat{\boldsymbol{\rho}}_\text{M}},
\end{equation}

\begin{equation}
    \setlength{\arraycolsep}{2pt}
    \medmuskip = 1mu
    \frac{\partial \varsigma^{^\text{\{BM,RM\}}}_2}{\partial \hat{\boldsymbol{\rho}}_\text{M}} = \frac{\partial \varsigma^{^\text{\{BM,RM\}}}_2}{\partial \theta'_{_\text{\{BM,RM\}}}}\frac{\partial\theta'_{_\text{\{BM,RM\}}}}{\partial \hat{\boldsymbol{\rho}}_\text{M}},
\end{equation}
\begin{equation}
    \setlength{\arraycolsep}{2pt}
    \medmuskip = 1mu
    \frac{\partial \omega_\text{\{BM,BRM\}}}{\partial \hat{\boldsymbol{\rho}}_\text{M}} = \frac{-2 \pi f_\text{sc} }{d_\text{\{BM,RM\}}c} \mathbf{I}_3\boldsymbol{\chi}_\text{\{BM,RM\}},
\end{equation}
\begin{equation}
    \setlength{\arraycolsep}{2pt}
    \medmuskip = 1mu
    \frac{\partial b_\text{BM}}{\partial \hat{\boldsymbol{\rho}}_\text{M}} = \frac{-c\mathbf{I}_3\boldsymbol{\chi}_\text{BM}}{4\pi f (d_\text{BM})^{3}} ,
    \frac{\partial b_\text{BRM}}{\partial \hat{\boldsymbol{\rho}}_\text{M}} = \frac{-b_\text{BR}c\mathbf{I}_3\boldsymbol{\chi}_\text{RM}}{4\pi f (d_\text{RM})^{3}} ,
\end{equation}
\begin{equation}
    \setlength{\arraycolsep}{2pt}
    \medmuskip = 1mu
    \frac{\partial \psi^{^\text{\{BM,RM\}}}_1}{\partial \textbf{r}_\text{M}} = \frac{\partial\psi^{^\text{\{BM,RM\}}}_1}{\partial \theta'_{_\text{\{MB,MR\}}}} \frac{\partial\theta'_{_\text{\{MB,MR\}}}}{\partial \textbf{r}_\text{M}} - \frac{\partial\psi^{^\text{\{BM,RM\}}}_1}{\partial \phi'_{_\text{\{MB,MR\}}}} \frac{\partial\phi'_{_\text{\{MB,MR\}}}}{\partial \textbf{r}_\text{M}},
\end{equation}
\begin{equation}
    \setlength{\arraycolsep}{2pt}
    \medmuskip = 1mu
    \frac{\partial \psi^{^\text{\{BM,RM\}}}_2}{\partial \textbf{r}_\text{M}} = \frac{\partial \psi^{^\text{\{BM,RM\}}}_2}{\partial \theta'_{_\text{\{MB,MR\}}}}\frac{\partial\theta'_{_\text{\{MB,MR\}}}}{\partial \textbf{r}_\text{M}}, \frac{\partial \varsigma^{^\text{\{BM,RM\}}}_1}{\partial \textbf{r}_\text{M}} =
    \frac{\partial \varsigma^{^\text{\{BM,RM\}}}_2}{\partial \textbf{r}_\text{M}} = 0,
\end{equation}
\begin{equation}
    \setlength{\arraycolsep}{2pt}
    \medmuskip = 1mu
    \frac{\partial \omega_\text{\{BM,BRM\}}}{\partial \textbf{r}_\text{M}} =
    \frac{\partial b_\text{BM}}{\partial \textbf{r}_\text{M}} =
    \frac{\partial b_\text{BRM}}{\partial \textbf{r}_\text{M}} = 0.
\end{equation}

\bibliographystyle{IEEEtran}
\mybibliography

\end{document}